\documentclass{iaus}
\usepackage{graphicx}

\title[Observational Study of IRAS 19312+1950] 
{Molecular Line Observations of the SiO Maser Source IRAS 19312+1950}

\author[J. Nakashima et al.]   
{J. Nakashima$^1$, S. Deguchi$^2$, H. Imai$^3$ \and A. Kemball$^4$}

\affiliation{$^1$Department of Physics, University of Hong Kong, Pokfulam Rd., Hong Kong
 \break email: junichi@hku.hk \\[\affilskip]
$^2$Nobeyama Radio Observatory, National Astronomical Observatory of Japan \\[\affilskip]
$^3$Department of Physics, Kagoshima University \\[\affilskip]
$^3$NCSA, University of Illinois at Urbana-Champaign}

\pubyear{2008}
\volume{251}  
\pagerange{001--002}
\date{version 02/28/2008}
\jname{Organic Matter in Space}
\editors{S. Kwok \& S. Sandford, eds.}
\begin{document}

\maketitle

\begin{abstract}
IRAS~19312+1950 is a unique SiO maser source, exhibiting a rich set of molecular radio lines, although SiO maser sources are usually identified as oxygen-rich evolved stars, in which chemistry is relatively simple comparing with carbon-rich environments. The rich chemistry of IRAS~19312+1950 has raised a problem in circumstellar chemistry if this object is really an oxygen-rich evolved star, but its evolutional status is still controversial. In this paper, we briefly review the previous observations of IRAS~19312+1950, as well as presenting preliminary results of recent VLBI observations in maser lines. PDF file of the poster is available from http://www.geocities.jp/nakashima\_junichi/
\end{abstract}

\firstsection 
\section{Introduction}
SiO maser sources are usually identified as oxygen-rich (O-rich) evolved stars, and it has been long thought that O-rich evolved stars do not significantly contribute to the enrichment of organic matter in the universe unlike carbon stars do. However, recent observations of IRAS~19312+1950, which is an SiO maser source (Nakashima \& Deguchi 2000), suggest that O-rich stars are non-negligible in the chemical evolution of the universe, because a rich set of molecular lines are detectable toward IRAS~19312+1950. If IRAS~19312+1950 is really an O-rich evolved star, its chemical properties are quite suggestive in terms of the interstellar chemistry, but we must securely identify its evolved-star status before going to further astrophysical/astrochemical interpretation, because some observational properties of this object are not consistent with the evolved-star status.

\section{Previous Observations}
The extended infrared nebulosity of IRAS~19312+1950 was first realized in the 2MASS images by Nakashima \& Deguchi (2000), and soon after that a more fine near-infrared image was obtained by UH 2.2~m/SIRIUS. The SIRIUS imaging revealed that the envelope of IRAS~19312+1950 exhibits a point symmetric structure. 

\begin{figure}
\centerline{
\scalebox{0.9}{%
\includegraphics{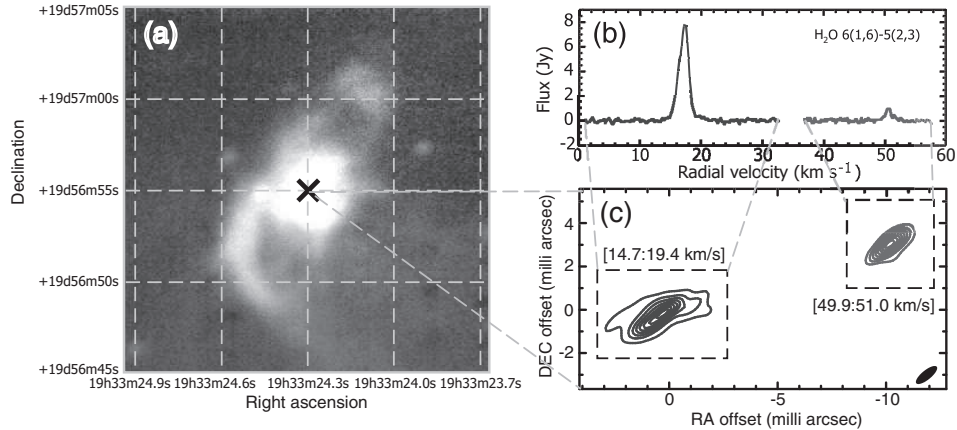}%
}
}
\caption{{\bf (a)}: Near-infrared H-band image of IRAS-19312+1950 taken by SUBARU/CIAO (Murakawa et al.~2007). {\bf (b)}: VLBA total flux profile of the H$_2$O maser line (6$_{1,6}$--5$_{2,3}$). {\bf (c)}: VLBA total intensity map of the red- and blue-shifted components of the H$_2$O maser line.}\label{fig:ratio-color}
\end{figure}

Deguchi et al.~(2004) and Nakashima et al.~(2004) searched molecular millimeter lines toward IRAS~19312+1950 using the NRO 45~m telescope. Even though the molecular lines search was not unbiased survey (i.e., selected lines were observed), they detected 22 different molecular species. Nakashima \& Deguchi (2005) made interferometric observations with BIMA in the bright molecular lines. They concluded that the properties of the broad component seen in the line profiles are explained by an expanding sphere model. 

\section{VLBA \& MERLIN Observations in Maser Lines}
Maser emission detected from IRAS~19312+1950 enables us to investigate the kinematics in the vicinity to the central star if we use the VLBI technique. We have recently made VLBA observations in the SiO ($J=1$--0, $v=1$ and 2) and H$_2$O (6$_{1,6}$--5$_{2,3}$) maser lines. In addition, we have made MERLIN observations in the H$_2$O (22 GHz) and OH (1612 MHz) maser lines. In the H$_2$O observations, we clearly detected a double peak in the total flux line profile, and confirmed that the red- and blue-shifted components are seated in spatially separated regions (see, Figure~1). 

\section{Discussion}
Our current interpretation of IRAS~19312+1950 is a rare AGB or post-AGB star embedded in a small dark cloud, even though we have no clear idea about why an AGB/post-AGB star is embedded in a small dark cloud. A recent model calculation by Murakawa et al.~(2007) have also suggested that the mass loss rate of the central star is consistent with that of an AGB star, but they have also withheld the discussion about the origin of the ambient material. To securely confirm the evolutional status of this object, presumably, observations listed below will be a key: (1) monitoring observations in maser lines with VLBI technique, (2) measuring a $^{12}$C/$^{13}$C ratio through, for example, radio observations in the H$^{12}$CO$^{+}$ and H$^{13}$CO$^{+}$ lines.


\end{document}